\documentclass[a4paper,11pt]{article}
\usepackage{jcappub} 
\usepackage{aas_macros}
\usepackage[utf8]{inputenc}
\usepackage[margin=0.80in]{geometry}
\usepackage{hyperref}
\usepackage{amsmath}
\usepackage[most]{tcolorbox}

\usepackage{enumitem}
\usepackage{subfigure}
\usepackage{booktabs}
\usepackage{caption}
\usepackage{graphicx}
\usepackage{multirow}
\usepackage{wrapfig}
\usepackage[capitalise,noabbrev]{cleveref}

\newcommand{\rmd}{\mathrm{d}}
\begin{document}
\title{Cosmology and Source Redshift Distributions from Combining Radio Weak Lensing with CMB Lensing}
\author[1,2]{A. Kalaja,}
\affiliation[1]{Van Swinderen Institute for Particle Physics and Gravity, University of
Groningen, Nijenborgh 4, 9747 AG Groningen, The Netherlands}
\affiliation[2]{Center for Computational Astrophysics, Flatiron Institute, 162 5th Avenue, New York, NY 10010, USA}
\author[3]{I. Harrison,}
\affiliation[3]{School of Physics and Astronomy, Cardiff University, The Parade, Cardiff, Wales CF24 3AA, UK}
\author[4,5]{William~R.~Coulton}
\affiliation[4]{Kavli Institute for Cosmology Cambridge, Madingley Road, Cambridge CB3 0HA, UK}
\affiliation[5]{DAMTP, Centre for Mathematical Sciences, University of Cambridge, Wilberforce Road, Cambridge CB3 OWA, UK}

\emailAdd{harrisoni@cardiff.ac.uk}
\emailAdd{wrc27@cam.ac.uk} 


\abstract{
Measurements of weak gravitational lensing using the cosmic microwave background and the shapes of galaxies have refined our understanding of the late-time history of the Universe. While optical surveys have been the primary source for cosmic shear measurements, radio continuum surveys offer a promising avenue. Relevant radio sources, principally star-forming galaxies, have populations with higher mean redshifts and are less affected by dust extinction compared to optical sources. We focus on the future mid frequency SKA radio telescope and explore the cross-correlation between radio cosmic shear and CMB lensing convergence ($\gamma_\mathrm{R}\times \kappa_\mathrm{CMB}$). We investigate its potential in constraining the redshift distribution of radio galaxy samples and improving cosmological parameter constraints, including the neutrino sector. Using simulations of the first phase of the SKA and the Simons Observatory as a CMB experiment, we show how this $\gamma_\mathrm{R}\times \kappa_\mathrm{CMB}$ cross-correlation can provide $\sim1 - 10\%$ calibration of the overall radio source redshift distribution, which in turn can significantly tighten otherwise degenerate measurements of radio galaxy bias. For the case of the next-generation full SKA, we find that the cross-correlation becomes more powerful than the equivalent with a \textit{Euclid}-like survey, with constraints $30\%$ tighter on $\Lambda$CDM parameters and narrower bounds on sum of neutrino masses at the level of $\sim 24\%$. These constraints are also driven by higher redshifts and larger scales than other galaxy-CMB cross-correlations, potentially shedding light on different physical models. Our findings demonstrate the potential of radio weak lensing in improving constraints, and establish the groundwork for future joint analyses of CMB experiments and radio continuum surveys.}

\maketitle
\section{Introduction}
\label{sec:intro}
In the past decades, weak gravitational lensing by the large-scale structure (LSS) of the universe has been established as a robust probe of late-time cosmology~\cite{lewis:weak_lensing_review,kilbinger:cosmic_shear}.
Lensing leaves an imprint in both the CMB and the images of galaxies. In the former, the distribution of matter on large scales alters the path of CMB photons as they travel from the last scattering surface to our detectors and thereby distorts the CMB anisotropies~\citep[see e.g.,][for a review]{lewis:weak_lensing_review}.  In the latter, the distribution of matter alters the path of photons from the galaxies leading to changes in the size, shapes and observed positions of the galaxies \citep[see e.g,][for a review]{Kilbinger_2015,Bartelmann_2017}. 

In the CMB, lensing induces a distinctive non-Gaussian signal upon the otherwise highly Gaussian primary temperature and polarization anisotropies. Isolating these anisotropies allows us to reconstruct the integrated line-of-sight mass at its origin, referred to as lensing convergence (denoted with $\kappa$).
Various CMB experiments have measured the power spectrum of $\kappa$~\cite{das:detection_lens_act,das:detection_lens_act2,planck13:gravitational_lensing,ade:polarbear_lens_detection,spt:lensing_measurement,story:spt_lens_detection2,bk_lens_detection,planck18:gravitational_lensing,qu:act_lensing23} and used it to constrain the dark energy equation of state and the sum of the neutrino masses (see e.g. Refs.~\cite{planck13:gravitational_lensing,battye:neutrino_mass_cmb_lensing}).

For galaxies, lensing leads to coherent distortions in the shapes of galaxies, typically called cosmic shear and represented as $\gamma$, that can be measured statistically. Measurements of galaxy shape correlations between different redshifts have led to competitive bounds on the expansion history of the Universe and the growth of cosmic structures (for example, see Ref.~\cite{hikage:subaru,2022PhRvD.105b3515S,2021A&A...645A.104A}). 

Up to now, observations of cosmic shear have relied to great extent on optical surveys due to the large number density of background galaxies observed. A promising new approach for cosmic shear measurements is represented by extragalactic radio sources~\cite{brown:ska_weak_lensing,2020PASA...37....7S}, which are dominated by two main galaxy populations~\cite{condon:radio_gal}: active galactic nuclei (AGN) and star-forming galaxies (SFGs). At radio frequencies, galactic extinction is less effective, allowing a larger sky coverage than optical surveys; source redshift distributions are expected to have a higher redshift tail contribution; cross-correlations may be expected to mitigate additive and multiplicative systematics \cite{harrison:skaforecasts,camera:skaforecasts}; and polarisation information may be available to lessen the impact of shape noise and intrinsic alignments \cite{2011MNRAS.410.2057B, Whittaker:2015fma}. 

The cosmological applications of radio data have been limited by the relatively low number density of radio sources reached in large scale radio surveys. Nonetheless, there exist a number of radio continuum surveys which approach the necessary samples, including the LOw Frequency ARray (LOFAR,~\cite{vanhaarlem:lofar}), the Meer
Karoo Array Telescope (MeerKAT,~\cite{jonas:meerkat}), the Australian Square Kilometre Array Pathfinder (ASKAP,~\cite{johnston:askap}), the Giant Metrewave Radio Telescope (GMRT,~\cite{swarup:gmrt}), and surveys, such as the NRAO VLA Sky Survey (NVSS,~\cite{condon:nvss}), the Faint
Images of Radio Sky at Twenty centimetres (FIRST) survey~\cite{becker:first} and the TIFR GMRT Sky Survey (TGSS-ADR,~\cite{intema:gmrt}). Moreover, cosmic shear has been detected with a 3.6$\sigma$ statistical significance using the FIRST survey~\cite{chang_radio_weak_lensing} and in galaxy-galaxy and cluster lensing cross-correlations between FIRST and SDSS \cite{demetroullas:ggl, demetroullas:sdss-first}. There are also ongoing efforts with the combination of JVLA and \emph{e}-MERLIN telescopes to detect the lensing signal as part of the SuperCLASS survey \cite{2020MNRAS.495.1706B, harrison:super_class}.


The first phase of the SKA, in particular the mid-frequency telescope being built in South Africa, will be capable of resolving an adequate number density of high redshift radio sources to perform a cosmologically informative weak lensing survey, offering a complementary approach to optical surveys \cite{harrison:skaforecasts, bonaldi:sims, camera:skaforecasts}. The SKA is expected to be built in two phases: SKA phase 1 (SKA-1), under construction and scheduled to be complete in 2030, and the full SKA (SKA-2), which will be active in the following decade.

In this work, we focus on both SKA-1 and SKA-2's mid frequency instrument as an example of a radio continuum telescope. SKA will access a larger radio source population than previous telescopes, with the dominant component expected to be formed by SFGs. Such galaxies are expected to be characterized by a long-tailed source redshift distributions, which extends the typical redshifts probed by optical surveys, as shown in detail in Ref.~\cite{brown:ska_weak_lensing}. Forecasts presented in Ref.~\cite{harrison:skaforecasts} for weak lensing experiments involving SKA show that the inclusion of such high-$z$ sources provides competitive constraints both from the radio waveband alone, and in cross-correlation with optical surveys.

One avenue that has not been explored yet is the cross-correlation of the cosmic shear from radio sources with the CMB lensing signal\footnote{The cross-correlation between radio sources and CMB lensing convergence has been used to address several aspects of both surveys, e.g. radio galaxy bias in Ref.~\cite{allison:act_radio_bias} and de-lensing in Ref.~\cite{namikawa:radio_delensing}. However, there is no work that explores the cosmic shear and lensing convergence cross-correlations for these calibrations (\cite{2015aska.confE..20K} forecasts the detection of the power spectrum only).}, hereafter $\gamma_\mathrm{R}\times \kappa_\mathrm{CMB}$. 
Indeed, as with optical, CMB and radio surveys depend on different systematics, meaning additive systematics are removed and multiplicative systematics can be self-calibrated out.

Moreover, cross-correlations with optical surveys are limited by the fact that sources are typically located at $z\sim 1$, while CMB lensing is most sensitive to structures at high redshifts ($z \simeq 1 - 5$). However, SKA will probe the galaxy population at higher redshifts than optical surveys, making $\gamma_\mathrm{R}\times \kappa_\mathrm{CMB}$ correlations potentially more informative than the optical counterpart. 

In this work, we forecast the impact of the $\gamma_\mathrm{R}\times \kappa_\mathrm{CMB}$ cross-correlation on two possible cosmological applications. We consider SKA for the radio experiment, Simons Observatory (SO) for the CMB lensing experiment, and show relative results with respect to a \emph{Euclid}-like optical weak lensing experiment.

First, we investigate its potential in constraining the redshift distribution of radio star-forming galaxies usable for weak lensing. A complication of radio weak lensing is the absence of reliable information about the redshifts of the sources. A common approach to this issue is by matching radio with the optical data~\cite{lindsay:bias_radio, hale:agn_bias_clustering, siewert:lofar}. Here, we exploit the extended overlap between the long-tail distribution of radio emission with CMB lensing to improve constraints on uncertainty parameters of the galaxy redshift distribution. This effectively uses the known redshift of the CMB lensing map to calibrate the unknown redshift distribution of the cosmic shear sources \cite{2020JCAP...12..001S}.

Secondly, we examine the extent to which $\gamma_\mathrm{R}\times \kappa_\mathrm{CMB}$ can improve constraints on cosmological parameters compared to an optical survey. We perform a Fisher matrix analysis for the current concordance $\Lambda$CDM model and explore the neutrino sector, in particular the sum of neutrino masses, $\sum m_\nu$.  For $\Lambda$CDM parameters we find cross-correlation between CMB lensing with radio cosmic shear from the full SKA can improve constraints by $\sim \mathcal{O}(30\%)$ compared to the optical cosmic shear cross-correlation case, where the exact number depends on the parameters considered. 

Massive neutrinos impact the total energy density of the Universe, $\Omega_\mathrm{m}$, and affect the formation and evolution of large-scale structure. Based on current observations, massive neutrinos are considered relativistic in the early Universe, whereas today their velocities have redshifted sufficiently that they contribute to $\Omega_\mathrm{m}$ as non-relativistic matter~\cite{lesgourgues:neutrinos2,lesgourgues:neutrinos1}. Depending on their masses, the presence of massive neutrinos is reflected in the position and the amplitude of the peaks in the CMB power spectra, due to changes in the angular distance of the sound horizon at recombination, and through the late integrated Sachs-Wolfe (ISW) effect. At the same time, due to their small masses, neutrinos suppress matter clustering at large and small scales. Because of this, their presence also affects the gravitational lensing of the CMB, allowing CMB observations to constrain $\sum m_\nu$. The most recent constraint on the sum of neutrino masses comes from the Atacama Cosmology Telescope (ACT), with $\sum m_\nu< 0.12 \;\mathrm{eV}$  (95\% confidence level)~\cite{act:neutrino_recent}.
Here, we show that $\gamma_\mathrm{R}\times \kappa_\mathrm{CMB}$ can potentially tighten the limit on $\sum m_\nu$ by $\mathcal{O}(20\%)$ with respect to the cross-correlation between SO and the \textit{Euclid}-like survey.

In our analysis we do not include uncertainties related to, e.g., galaxy intrinsic alignments and modelling of the non-linear matter clustering. We therefore expect that our absolute bounds are optimistic and instead focus on the relative constraining power between the different experiments and scenarios.


The rest of the paper is structured as follows: in Sec.~\ref{sec:methodology}, we introduce the theoretical framework and describe the experiments considered. In Sec.~\ref{sec:mcmc_dndz} we provide new parameterised fits for redshift distributions of simulated radio populations using our forecast machinery. In Sec.~\ref{sec:cosmo_application} we show our main results: on calibrating radio redshift distributions using the first phase of the SKA, and on both this information and $\Lambda$CDM and neutrino cosmological parameters for the full SKA. In Sec.~\ref{sec:conc} we draw conclusions and give an insight on future perspectives.
\paragraph*{{Notation and conventions in this chapter}} 
In our numerical computations, we consider a flat $\Lambda$CDM cosmology, with cosmological parameters in accordance with the latest \textit{Planck} results~\cite{planck2018:cosmological_parameters}, summarized in Tab.~\ref{tab:cosmology_rwl}. 
\begin{table}[t]
    \centering
    \begin{tabular}{c  c}
     \toprule
     \textbf{Parameter} & \textbf{Value}\\
     \midrule
        $H_0$ & $67.32\,\text{km\,s}^{-1}\,\text{Mpc}^{-1}$ \\
        $\Omega_\mathrm{b} h^2$ & $0.02238$ \\
        $\sum m_\nu$ & $0.06\,\text{eV}$\\
        $\Omega_\mathrm{k}$ & $0$ \\
        $\Omega_\mathrm{cdm} h^2$ & $0.12010$\\
        $\tau$ & $0.0543$\\
        $n_\mathrm{s}$ & $0.9660$\\
        $A_\mathrm{s}$ & $2.1005\times 10^{-9}$\\
    \bottomrule
    \end{tabular}
    \caption[Best-fit Planck parameters used in our numerical computations.]{Best-fit cosmological parameters from the \emph{Planck} experiment (specifically, Tab.~1 of Ref.~\cite{planck2018:cosmological_parameters} with $TT,TE,EE\text{+low$E$+lensing+BAO}$) used in our numerical computations. These are the fiducial cosmology parameters we assume and calculate our predicted constraining power around.}
    \label{tab:cosmology_rwl}
\end{table}
\section{Methodology}
\label{sec:methodology}
CMB lensing and cosmic shear trace the same underlying matter distribution at different redshifts, therefore their cross-correlation represents a robust observable to test the evolution of structure formation and investigate the nature of the dark components of the universe. 
Given two observables $(A,\,B)$, the (non-tomographic) cross-correlated power spectrum is given by
\begin{equation}
C^{AB}(\ell) = 8\pi^2 \int \frac{\rmd\chi}{\ell^3}\chi\, W_A(\chi,\chi_*)W_B(\chi,\chi_*)\mathcal{P}_\Psi\left(k=\frac{\ell+1/2}{\chi};\tau_0,\chi\right),
\end{equation}
where we used the Limber approximation~\cite{limber:approximation}. The power spectra are sensitive to the values of cosmological parameters via $\mathcal{P}_\Psi$. For example in the linear regime  
\begin{equation}
    \frac{2\pi^2}{k^2}\mathcal{P}_\Psi = \left(\frac{3}{2}\frac{\Omega_\mathrm{m} H^2_0}{ k^2a}\right)^2 P_\delta = \left(\frac{3}{2}\frac{\Omega_\mathrm{m} H^2_0}{ k^2a}\frac{T(k)D_+(t)}{D_+(t_i)} \right)^2 P_\zeta,
\end{equation}
where $P_\delta$ is the matter power spectrum $P_\zeta$ is the curvature perturbation produced in the early Universe, $D_+$ is the linear growth factor and $T(k)$ is the matter transfer function.
\begin{table}[h!]
\renewcommand{\arraystretch}{1.2} 
\centering
\begin{tabular}{ccccccc}
\toprule
Experiment           & $f_\mathrm{sky}$ & $n_\mathrm{gal}$  ($\mathrm{arcmin}^{-2}$) & $z_\mathrm{m}$ & $\alpha$   & $\beta$ & $\gamma$ \\
\midrule
SKA-1                & 0.12       & 2.7  & 1.1            & $\sqrt{2}$ & 2       & 1.25   \\
SKA-2                & 0.7       & 10  & 1.3            & $\sqrt{2}$ & 2       & 1.25    \\
\textit{Euclid}-like & 0.36      & 30  & 0.9            & $\sqrt{2}$ & 2       & 1.5     \\  
\bottomrule
\end{tabular}
\caption[Parameters used for the experiments considered in this work.]{Fiducial parameters in the galaxy redshift distribution function \cref{{eq:dndz_def_1}} used for the representative experiments considered in this work, as taken from Table 1. of \cite{harrison:skaforecasts}. We further refine the values of these parameters by fitting them with our cross-correlation observables directly to the galaxy number densities in the T-RECS \cite{bonaldi:t_recs} and SKADS \cite{wilman:skads} simulations.}
\label{tab:exp_dndz}
\end{table}
\begin{figure}[h!]
    \centering
    \includegraphics[width=0.8\linewidth]{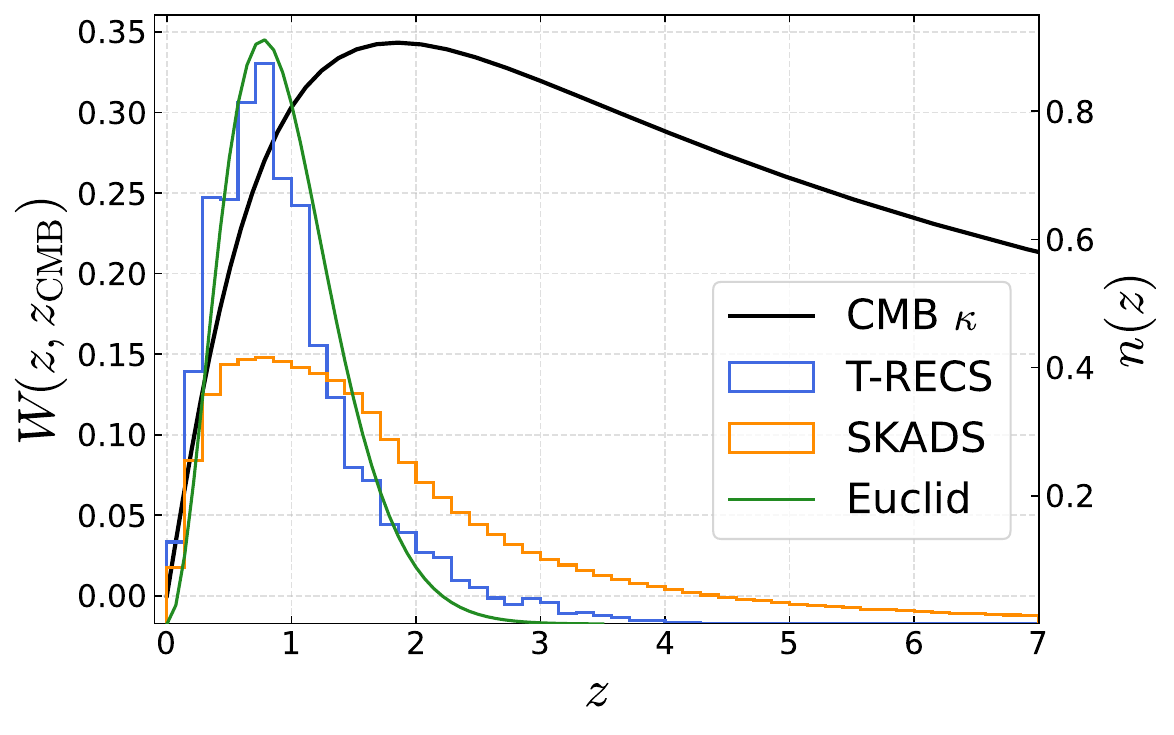}
    \caption[T-RECS, SKADS and \textit{Euclid} redshift galaxy distribution, and CMB lens efficiency.]{The plot shows the redshift galaxy distributions from T-RECS, SKADS and \textit{Euclid}-like experiments we consider, and the CMB lensing efficiency in Eq.~\eqref{eq:w_cmb}. The latter has been normalized such that the area under the curve is equal to one.}
    \label{fig:angsize_dndz_trecs_skads_cmb}
\end{figure}
We include non-linear corrections to the matter power spectrum, modeled with \texttt{halofit}~\cite{bird:halofit,takahashi:halofit}. In this context, the lens efficiency can either be the CMB lensing kernel given by
\begin{equation}\label{eq:w_cmb}
    W_\mathrm{CMB}(\chi) = \frac{\chi_\mathrm{CMB} - \chi}{\chi_\mathrm{CMB} \chi}.
\end{equation}
or the cosmic shear kernel 
\begin{equation}
    W_\mathrm{gal}(\chi) = \int_{\chi}^{\chi_*}\rmd\chi' n(\chi') \frac{\chi_* - \chi}{\chi_* \chi}.
\end{equation}
Here, $\chi_*$ is the comoving distance of the source, with $\chi_\mathrm{CMB}$ corresponding to the distance from the last scattering surface, and the redshift distribution of galaxies, given by $\frac{\mathrm{d}n}{\mathrm{d}z}$. Here we consider two different possible parameterisations of the $\frac{\mathrm{d}n}{\mathrm{d}z}$ (which we will also fit to two different simulation models of the radio source populations, as described in \cref{subsec:experiments}). We look at how we are able to constrain their parameters with our cross-correlation observable, and in turn how marginalising over them as nuisance parameters affects constraints on cosmological parameters. The first $\frac{\mathrm{d}n}{\mathrm{d}z}$ we consider is the popular mixed power law-exponential model (often referred to as the `Smail distribution'):
\begin{equation}\label{eq:dndz_def_1}
    \frac{\mathrm{d}n}{\mathrm{d}z} = z^\beta e^{-(z/z_0)^\gamma}, \quad z_0 = z_\mathrm{m}/\alpha,
\end{equation}
normalized such that:\footnote{We use the fact that $n(\chi)\rmd \chi = n(z)\rmd z$.}
\begin{equation}
    n(z) = \frac{1}{\bar{n}}\frac{\rmd n}{\rmd z},\quad \text{with}\quad \frac{1}{\bar{n}} = \int_{0}^{z_\star}\mathrm{d}z\,\frac{\mathrm{d}n}{\mathrm{d}z}.
\end{equation}
The free parameters \{$\beta$, $\gamma$, $z_\mathrm{m}$, $\alpha$\} depend on the specific characteristics of the experiment, with $z_\mathrm{m}$ the median redshift of sources. We report in Tab.~\ref{tab:exp_dndz} the values of these parameters for SKA and a \textit{Euclid}-like survey as were used in previous SKA cosmic shear forecasts \cite{harrison:skaforecasts,bonaldi:sims,camera:skaforecasts}. In particular, the values shown for SKA are parameters which reasonably match the redshift distributions for a weak lensing sample extracted from the SKADS S3-SEX \cite{wilman:skads} simulations of radio source populations (with modifications to match newer data as described in \cite{bonaldi:sims}). In the following \cref{sec:mcmc_dndz}, we perform updated explicit fits for values of $(\beta,\,\gamma)$ for SKADS and for a more recent suite of simulations, T-RECS \cite{bonaldi:t_recs}.

We also consider the $z_{\rm tail}$ model for $\frac{\mathrm{d}n}{\mathrm{d}z}$, which parameterises solely the redshift tail of the distribution and has previously been used in the cross-correlation between CMB lensing maps and galaxy \textit{clustering} \cite{alonso:crosscorr}, where it was found to be significantly degenerate with the linear galaxy bias $b$, an important nuisance parameter for studies involving galaxy clustering two-point functions. The $z_{\rm tail}$  model has the form:
\begin{equation}\label{eq:dndz_def_2}
    \frac{\mathrm{d}n}{\mathrm{d}z} = \frac{\left(z / z_0\right)^2}{1 + \left(z / z_0\right)^2} \frac{1}{ 1 + \left(z/z_{\rm tail}\right)^{\gamma_{\rm tail}}}, \quad z_0 = z_\mathrm{m}/\alpha,
\end{equation}
In common with previous approaches (and to give an example where only one additional parameter is added to characterise the redshift distribution) we only consider varying the $z_{\rm tail}$ parameter in this case, with the parameters $\lbrace \alpha, \gamma_{\rm tail} \rbrace$ kept fixed.


\subsection{Experiments considered}
\label{subsec:experiments}
The results presented here are based on simulations of the radio sky from SKADS and T-RECS, which are specifically designed to include the relevant source populations for wide field continuum SKA cosmology surveys. For each simulation, we fit parameterised models of the forms in \cref{eq:dndz_def_1,eq:dndz_def_2} and include their parameters in our Fisher matrix calculations to see how well they may be constrained.

SKADS uses a `semi-empirical' approach to simulate radio sources, where they are generated by sampling observed radio continuum luminosity functions for extragalactic populations. 
The simulations feature a sky area equivalent to\footnote{This is approximately the largest \textit{instantaneous} field of view of SKA.} $20\times 20 \, \mathrm{deg}^2$, and reach a maximum redshift of $z_\mathrm{max}=20$. The flux density limit is $10\,\mathrm{nJy}$ over the $151\, \mathrm{MHz}-18\, \mathrm{GHz} $ frequency range. The redshift distribution of sources is then tailored to match the luminosity function of different types of radio sources, which consists of `radio-loud' (RL) AGNs, and `radio-quiet' (RQ) AGNs and SFGs. The clustering properties of radio sources are also included in the simulations by assuming a model for their bias. We use the updates to the SKADS populations proposed by Ref.~\cite{bonaldi:sims} to match the most recent data. Specifically, we re-calibrate the overall number of SFGs and the angular size distribution of the sources.  

The more recent T-RECS suite of simulations spans a frequency range of $150\, \mathrm{MHz}-20\, \mathrm{GHz} $ and reproduces the most recent data in terms of number counts, luminosity functions and redshift distributions. It is organized in three tiers based on field of view and flux limit: ``deep'' ($1\, \mathrm{deg}^2$, $1\,\mathrm{nJy}$), ``medium'' ($25 \, \mathrm{deg}^2$, $10\,\mathrm{nJy}$), and ``wide'' ($400 \, \mathrm{deg}^2$, $100\,\mathrm{nJy}$). 
The radio sources consist of two main populations, i.e. AGNs and SFGs. T-RECS does not model explicitly RQ-AGNs, but their contribution to the overall flux can be approximated by considering them as part of the SFG population. The radio sources are assigned to the dark matter halos in the light-cone out to a redshift of $z_\mathrm{max} = 8$, while ensuring they match the luminosity functions and clustering properties of AGNs and SFGs.

SKADS predicts a higher fraction of SFGs at low redshifts and a more extended redshift tail than T-RECS, see e.~g. Fig.~\ref{fig:angsize_dndz_trecs_skads_cmb}. Given our interest in the cross-correlation with CMB lensing, we consider the $n(z)$ distributions up to $z = 7$ for both sets of simulations. However, it is known that there exist uncertainties associated with the astrophysical sources beyond $z \approx 3$~\cite{allison:act_radio_bias}.

For the $\gamma_\mathrm{R}\times \kappa_\mathrm{CMB}$ cross-correlation, we consider SO as the CMB experiment. For SO we use a simulated lensing noise curve for the baseline sensitivity and no foreground deprojection, as discussed in \cite{simons:forecasts}. Specifically we use the recommended noise curve  in the public SO noise models repository\footnote{\texttt{nlkk\_v3\_1\_0\_deproj0\_SENS1\_fsky0p4\_it\_lT30-3000\_lP30-5000.dat} from \url{https://github.com/simonsobs/so_noise_models/}}
Additionally, in order to get a sense of the relative constraining power of the $\gamma_\mathrm{R}\times \kappa_\mathrm{CMB}$, we include the cross-correlation of CMB lensing with cosmic shear from an optical survey following the specifications of the \textit{Euclid}-like survey in \cref{tab:exp_dndz}, referred to as $\gamma_\mathrm{O}\times \kappa_\mathrm{CMB}$.
\begin{figure}[h!]
    \centering
    \includegraphics[width=1\linewidth]{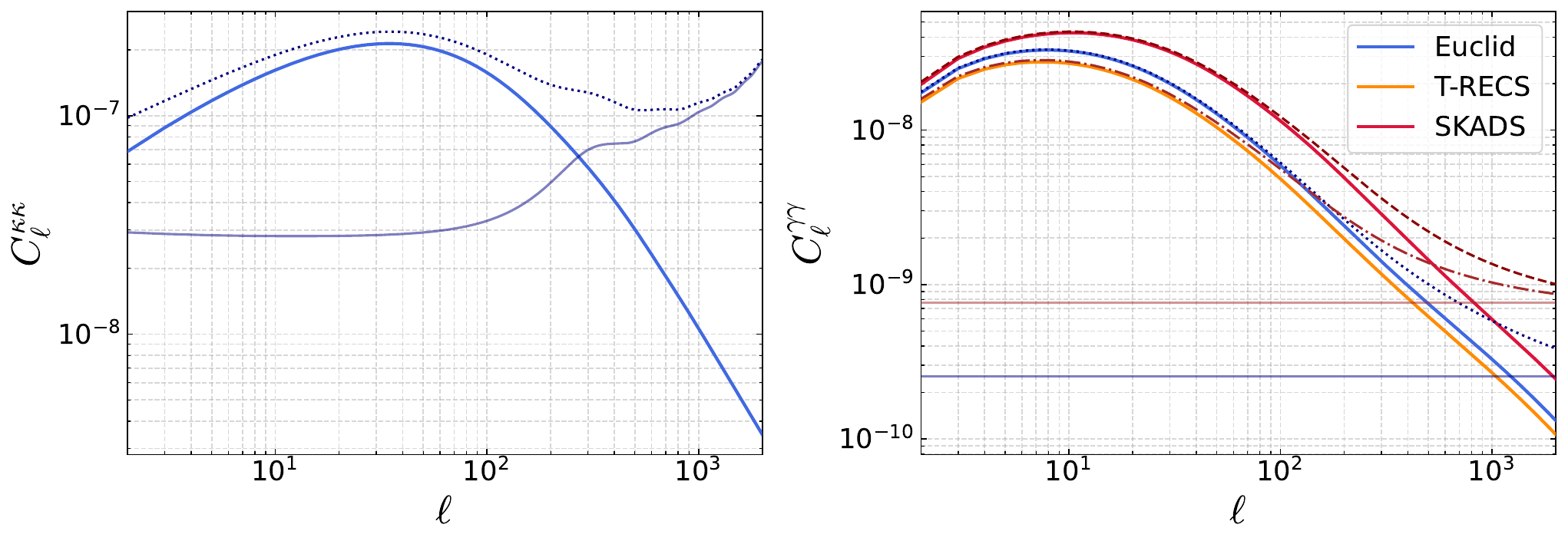}
    \caption[Power spectra and noise curves for SO, \textit{Euclid} and SKA-2 (T-RECS, SKADS).]{Power spectra and noise curves for SO, \textit{Euclid} and SKA-2 (T-RECS, SKADS). Solid lines refer to the cosmological power spectra; faded lines are the noise curves; the dashed, dashed, dotted and dotted lines are the corresponding total power spectra.}
    \label{fig:cls_noise}
\end{figure}
\subsection{Parameter estimation}\label{subsec:par_est}
In this work, we use two methods to infer parameters from our simulated data: a Markov Chain Monte Carlo (MCMC) and a Fisher matrix.
The analysis consists of two main steps:
\begin{itemize}
    \item First, we use an MCMC analysis with fixed true cosmological parameters to directly infer the parametrised $n(z)$ from the suites of simulations described in Subsec.~\ref{subsec:experiments}, to update the best-fitting values of $\{\beta_\mathrm{BF},\,\gamma_\mathrm{BF}\}$ parameters in Eq.~\eqref{eq:dndz_def_1}. 
    \item Secondly, we use these values $\{\beta_\mathrm{BF},\,\gamma_\mathrm{BF}\}$ in the  parametrization of $n(z)$ to compute the ``true'' power spectra, $C^{AB}_\ell$. We then perform a Fisher matrix forecast on \{$\Omega_\mathrm{cdm}$, $\Omega_\mathrm{b}$, $h$, $n_\mathrm{s}$, $A_\mathrm{s}$\} and the redshift distribution parameters. We compare the constraints across the different simulations and with the illustrative example for the optical survey $\gamma_\mathrm{O}\times \kappa_\mathrm{CMB}$. We apply the same analysis to an extended parameter space including the neutrino mass $\sum m_\nu$.
\end{itemize}
Assuming a Gaussian likelihood, $\mathcal{L}$ for the power spectra, the likelihood is given by
\begin{align}\label{eq:deviance}
-2 \ln \mathcal{L} = \sum_{\ell=\ell_\mathrm{min}}^{\ell_\mathrm{max}} \left(\boldsymbol{C}_\ell(\boldsymbol\vartheta) - \widetilde{\boldsymbol d}_{\ell}\right)  \boldsymbol{\Gamma}_{\ell}^{-1} \left(\boldsymbol{C}_{\ell}(\boldsymbol\vartheta) - \widetilde{\boldsymbol d}_{\ell}\right),
\end{align}
where $\boldsymbol\vartheta$ is the parameter-vector containing the parameters that we aim to constrain. The vectors $\boldsymbol{C}_\ell$ and $\widehat{\boldsymbol d}_{\ell}$ are respectively the fiducial power spectra and the data vector, which includes the `observed' auto- and cross-correlated spectra, $\hat{C}_\ell^{AB}$. Specifically:
\begin{equation}
\boldsymbol{C}_\ell = 
\begin{pmatrix}
C_\ell^{\kappa\kappa} \\
C_\ell^{\kappa \gamma} \\
C_\ell^{\gamma\gamma}
\end{pmatrix},\qquad \widehat{\boldsymbol{d}} = 
\begin{pmatrix}
\hat{C}_\ell^{\kappa\kappa} \\
\hat{C}_\ell^{\kappa \gamma} \\
\hat{C}_\ell^{\gamma\gamma}
\end{pmatrix}.
\end{equation}
A key element of the Gaussian likelihood is the covariance matrix, given by
\begin{align}
\boldsymbol{\Gamma}_{\ell} =  \frac{1}{(2\ell + 1)f^X_\mathrm{sky}} \begin{pmatrix}
2(\hat{C}_\ell^{\kappa \kappa,\,\rm tot})^2 & 2\hat{C}_\ell^{\kappa \kappa,\,\rm tot}\hat{C}_\ell^{\kappa \gamma,\,\rm tot} & 2(\hat{C}_\ell^{\kappa \gamma,\,\rm tot})^2 \\
2\hat{C}_\ell^{\kappa \kappa,\,\rm tot}\hat{C}_\ell^{\kappa \gamma,\,\rm tot} & (\hat{C}_\ell^{\kappa \gamma,\,\rm tot})^2 + \hat{C}_\ell^{\kappa \kappa,\,\rm tot}C_\ell^{\gamma\gamma,\,\rm tot} & 2\hat{C}_\ell^{\kappa \gamma,\,\rm tot}\hat{C}_\ell^{\gamma\gamma,\,\rm tot} \\
2(\hat{C}_\ell^{\kappa \gamma,\,\rm tot})^2 & 2\hat{C}_\ell^{\kappa \gamma,\,\rm tot}\hat{C}_\ell^{\gamma \gamma,\,\rm tot} & 2(\hat{C}_\ell^{\gamma \gamma,\,\rm tot})^2
\end{pmatrix},
\end{align}
where we included the noise spectra as
\begin{equation}
    \hat{C}^{AB,\,\rm tot}_\ell = \hat{C}^{AB}_\ell + N^{AB}_\ell.
\end{equation}
Specifically, we use the publicly available SO noise curves for $N^{\kappa\kappa}_\ell$, whereas the uncertainty on the shear spectrum depends on $n_\mathrm{gal}$, the number density of detected galaxies on the sky, and $\sigma^2_{\rm gal}$, the variance of the distribution of galaxy ellipticities (or ‘\textbf{shape noise}’)~\cite{hu:joint_gal_lens}
\begin{equation}
    N^{\gamma\gamma}_\ell = \frac{\sigma^2_{\rm gal}}{n_{\rm gal}},\quad \sigma_{\rm gal} = 0.3.
\end{equation}
Because the noise curves associated to SO and to SKA/\textit{Euclid} are uncorrelated, $N^{\kappa\gamma}_\ell = 0$. In \cref{fig:cls_noise} we show the respective noise curves alongside the fiducial cosmological signal, demonstrating their relative amplitude and angular scale dependence in both probes.

We sample the Likelihood using the \texttt{emcee} library~\cite{emcee}. We further assume a fraction of the sky $f^{\mathrm{SO}-\mathrm{SKA-1}}_\mathrm{sky} \approx 0.12$, $f^{\mathrm{SO}-\mathrm{SKA-2}}_\mathrm{sky} \approx 0.48$ and $f^{\mathrm{SO}-\mathrm{Eucl}}_\mathrm{sky} \approx 0.25$, corresponding to the overlap between the SO-LAT (Large Aperture Telescope) and, respectively, the SKA-1, SKA-2 and \textit{Euclid} sky coverage. 

Finally, assuming a Gaussian likelihood and using the same notation as above, the Fisher matrix is given by
\begin{equation}
    F_{ij} = \sum_{\ell=\ell_\mathrm{min}}^{\ell_\mathrm{max}}\frac{\partial \boldsymbol{C}_\ell(\boldsymbol\vartheta)}{\partial \boldsymbol{\vartheta}_i} \boldsymbol{\Gamma}_{\ell}^{-1} \frac{\partial \boldsymbol{C}_{\ell}(\boldsymbol\vartheta)}{\partial \boldsymbol{\vartheta}_j}.
\end{equation}

\section{Monte Carlo fitting of galaxy redshift distribution models}
\label{sec:mcmc_dndz}
As an intermediate step in our analysis we generate explicit fits for the redshift $\{\beta,\, \gamma\}$ parameters by Monte Carlo sampling from their posterior using the log-likelihood defined in Eq.~\eqref{eq:deviance}. We generate initial values for walkers in a ball around the $\{\beta,\, \gamma\}$ listed in Tab.~\ref{tab:exp_dndz}, and assume broad flat priors. Other cosmological and redshift parameters are kept fixed at the fiducial values shown in \cref{tab:exp_dndz,tab:cosmology_rwl}. This fitting step ensures we are using appropriate $n(z)$ in the subsequent analyses. Fitting the $\beta$ and $\gamma$ via the $C_\ell$ likelihood in this way both (as opposed to directly to the $n(z)$ histograms) accounts for the expected information content we have on the parameters, and checks that their posteriors are Gaussian and hence appropriate for approximating with Fisher matrices in the subsequent sections.

\begin{figure}
    \centering
    \includegraphics[width=0.9\linewidth]{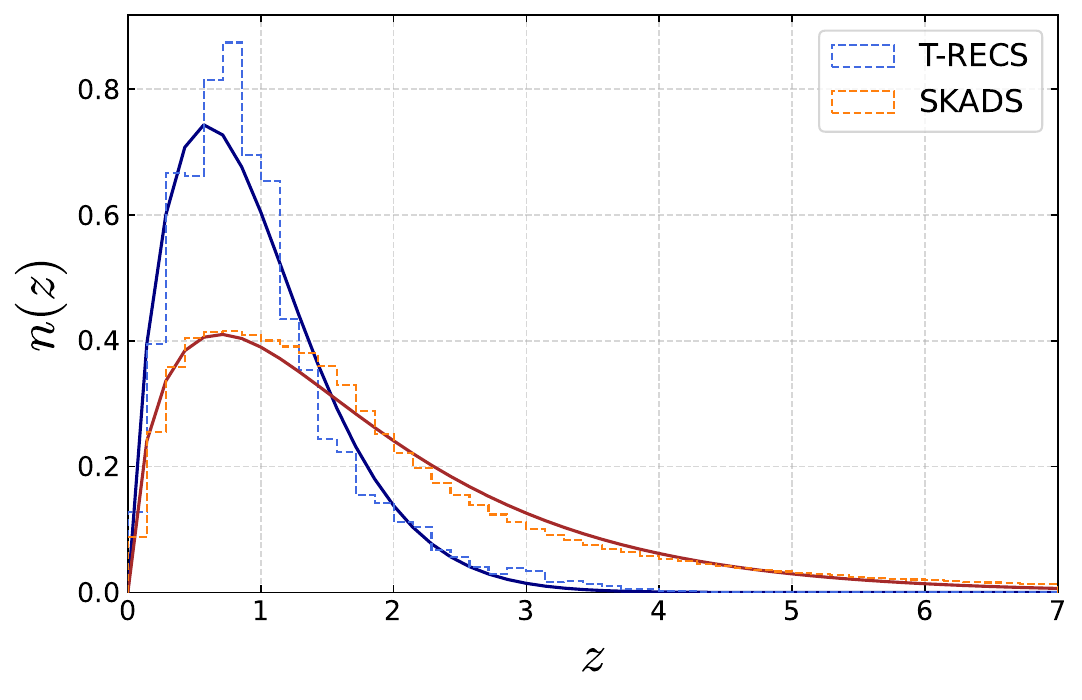}
    \caption[Best-fit galaxy distribution for T-RECS and SKADS.]{The plot shows the comparison between the galaxy distribution from the T-RECS and SKADS simulations (solid lines) and the fit obtained using the best-fit parameters shown in Tab.~\ref{tab:bf_beta_gamma} from the Monte Carlo fitting procedure described in \cref{sec:mcmc_dndz}.}
    \label{fig:bf_dndz}
\end{figure}

\begin{figure}[h!]
    \centering
        \includegraphics[width=0.9\linewidth]{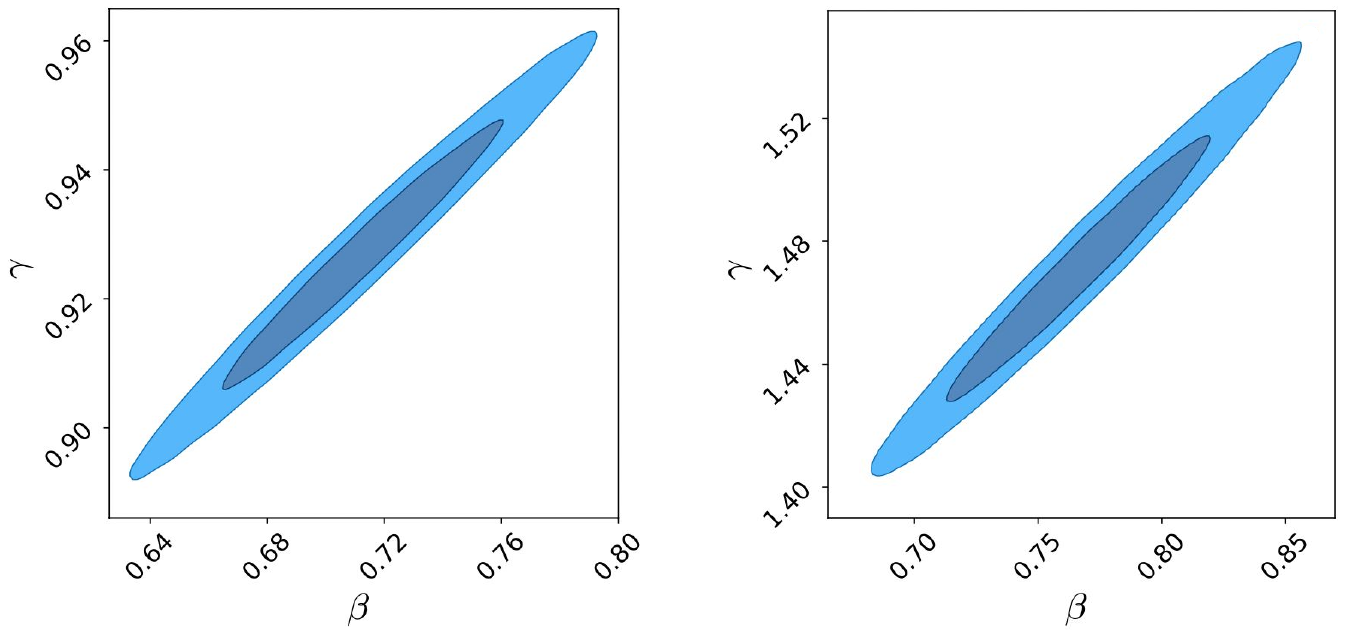} 
    \caption[Marginalized 68\% and 95\% parameter constraint contours for $\{\beta,\, \gamma\}.$]{Marginalized 68\% and 95\% parameter constraint contours obtained in \cref{sec:mcmc_dndz} for the T-RECS (\textit{left panel}) and SKADS (\textit{right panel}) simulations. This demonstrates how well the parameters governing the radio source redshift distribution ($\beta$ and $\gamma$) will be constrained with CMB cross correlations. }
    \label{fig:postdist_ska}
\end{figure}

The resulting posterior distributions for T-RECS and SKADS are reported in Fig.~\ref{fig:postdist_ska}, and the corresponding best-fit value can be found in Tab.~\ref{tab:bf_beta_gamma}. 
Fig.~\ref{fig:bf_dndz} shows the excellent agreement with the $n(z)$ from simulations.
\begin{table}[h!]
\renewcommand{\arraystretch}{1.2} 
\centering
\begin{tabular}{ccccc}
\toprule
Simulation         & $\beta$ & $\gamma$ & $\gamma_{\rm tail}$ & $z_{\rm tail}$ \\ \midrule
T-RECS                & $0.767^{+0.035}_{-0.034}$       &  $1.472\pm 0.028$ & 3.5 & 0.65 \\ \midrule
SKADS & $0.712\pm 0.031$      & $0.927^{+0.013}_{-0.014}$ & 2.5 & 0.74 \\  
\bottomrule
\end{tabular}
\caption[Best-fit parameters for $\{\beta,\,\gamma\}$ using T-RECS and SKADS.]{Best-fit parameters for the galaxy number density distributions Eq.~\eqref{eq:dndz_def_1} and Eq.~\eqref{eq:dndz_def_2} (with error bars from the Monte Carlo fitting) when using the T-RECS and SKADS simulations. We use these as the fiducial values for the number density model in our Fisher analysis in \cref{sec:cosmo_application}.}
\label{tab:bf_beta_gamma}
\end{table}
This exercise provides us with best-fitting values for the redshift distribution parameters, which we take forward into the Fisher analysis of constraining power in the next section. For the $z_{\rm tail}$ parameterisation \cref{eq:dndz_def_2} we directly minimise the log-likelihood to find initial values of the fits to the two distributions, with the resulting values show in \cref{tab:bf_beta_gamma}.

\section{Cosmological applications}
\label{sec:cosmo_application} 
Next, we use the results of Sec.~\ref{sec:mcmc_dndz} on the galaxy distribution to test the constraining power of the full $\gamma_\mathrm{R}\times \kappa_\mathrm{CMB}$ cross-correlation. We use the Fisher formalism described in Subsec.~\ref{subsec:par_est}, where we include auto- and cross-power spectra. In particular, we forecast the $1\sigma$ errors on the redshift parameters and the fiducial $\Lambda$CDM model set of parameters
\begin{equation*}
    \{\omega_\mathrm{cdm},\; \omega_\mathrm{b},\; h,\; n_\mathrm{s}, \;A_\mathrm{s}\},
\end{equation*}
where $\omega_\mathrm{x} = \Omega_\mathrm{x}h^2$. We fix $\sum m_\nu = 0.06$, with one massive and two massless neutrinos. We compare the results to the cross-correlation of CMB lensing with optical-based shear measurements with \textit{Euclid} specifications.

\subsection{Inference on Redshift Distributions with SKA-1 and SKA-2}
\begin{figure}[h!]
        \centering
        \includegraphics[width=0.5\linewidth]{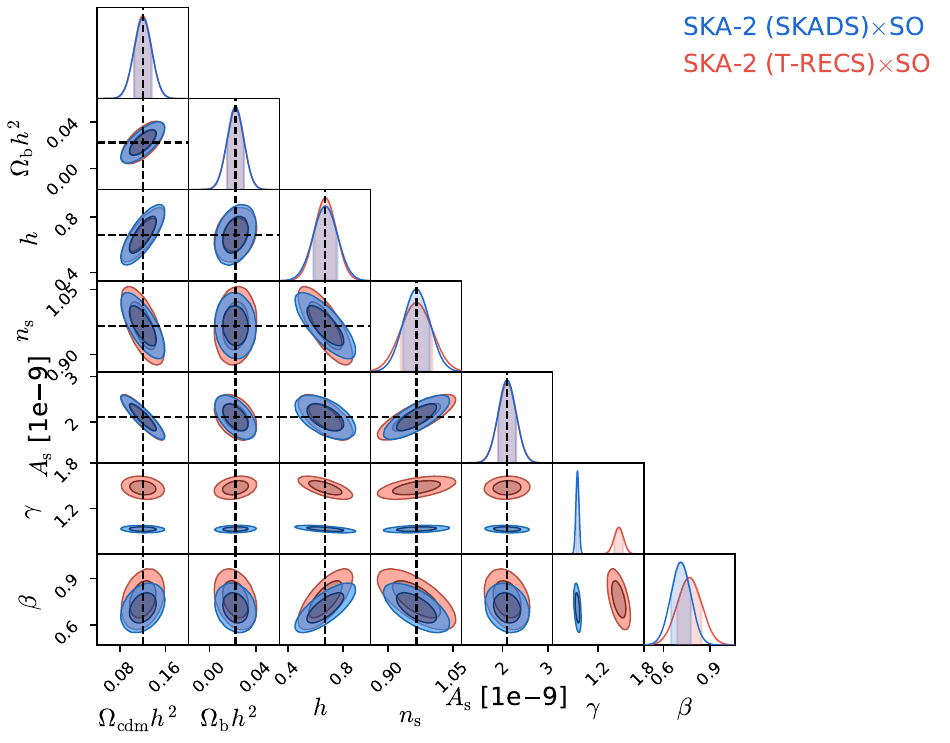}\includegraphics[width=0.5\linewidth]{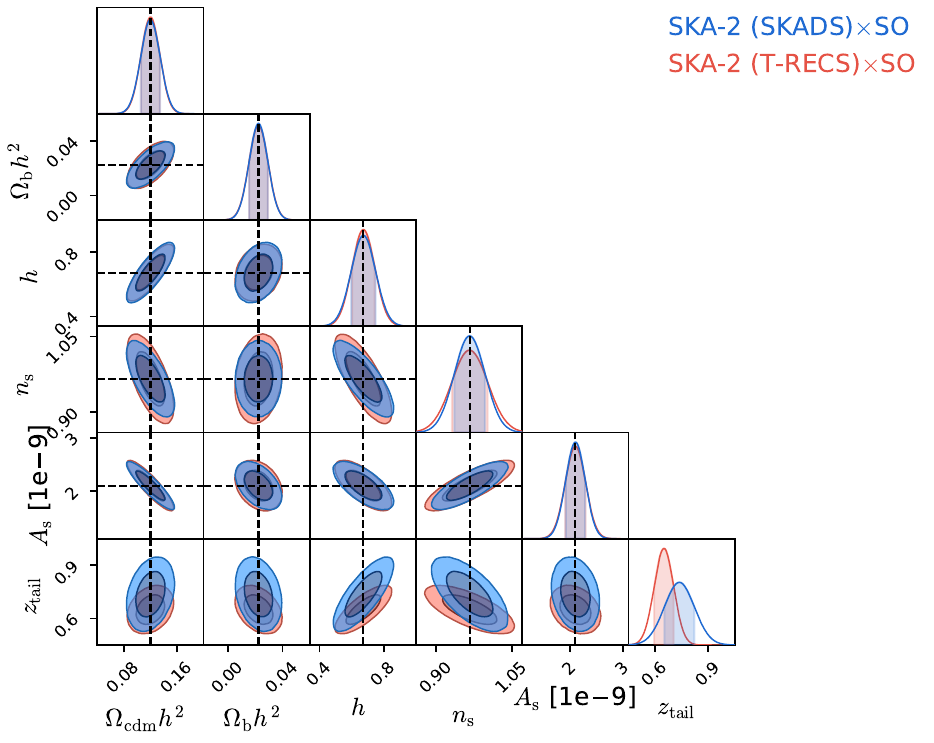}
        \caption{Fisher prediction of 68\% and 95\% contours on on the fiducial $\Lambda$CDM parameters from the full (auto- and cross-correlation) $\gamma\times\kappa_\mathrm{CMB}$ analysis with radio and optical surveys. The left figure corresponds to the Smail redshift distribution and the right to the $z_\mathrm{tail}$ model. Here we show the relative agreement across the two simulations and degeneracies between the cosmological and redshift parameters. We refer to \cref{fig:improvements} for comparisons of the relative constraining power to a \emph{Euclid}-like experiment.}
        \label{fig:zparm_fisher}
\end{figure}

We first consider the case where we may only be interested in redshift parameters for the radio surveys, particularly relevant for the SKA-1 era when the cosmological constraining power is relatively less than other experiments and where knowledge of the radio source redshift parameters may still be highly uncertain. In \cref{tab:zparm_fisher} and \cref{fig:zparm_fisher} we show the results on $\beta, \gamma$ and $z_{\rm tail}$ for SKA-1 in the different cases of both SKADS and T-RECS redshift distributions. We consider both cases where cosmological parameters are kept fixed (also representative of a strong prior relative to this data from e.g. \emph{Planck}+BAO) and where they are also varied and marginalised over (without any explicit external prior from another experiment). When including cosmological marginalisation uncertainties are typically in the tens of per-cent, and for the case of fixed cosmology in the one-to-ten per-cent range. These numbers compare favourably with the approach of inferring this information from galaxy clustering cross-correlations in current data, with \cite{alonso:crosscorr} obtaining uncertainties of $\sim25\%$ on $z_{\rm tail}$ at a fixed cosmology, driven in part by having to jointly infer the redshift parameter with the galaxy linear bias, which is not necessary for the lensing combination used here.
\begin{table}
    \renewcommand{\arraystretch}{1.2} 
    \centering
    \caption{The table reports the predicted fractional $1\sigma$ errors on redshift parameters when cosmological parameters are either fixed or marginalised, using the different radio experiments considered and for the different simulations of galaxy number densities.}
    \label{tab:model_params}
    \begin{tabular}{lccc}
        \hline
        Experiment & $\sigma_\beta$ & $\sigma_\gamma$ & $\sigma_{z_{\rm tail}}$\\
        \hline
        SKA-1 (T-RECS) & 14\% & 5.6\% & 1.4\% \\
        inc. cosmology marg. &  39\% & 15\% & 32\% \\
        \hline
        SKA-1 (SKADS) & 11\% &  3.3\% & 1.5\% \\
        inc. cosmology marg. & 33\% & 6.7\% & 42\% \\
        \hline
        SKA-2 (T-RECS) & 4.5\% & 1.9\% & 0.31\% \\
        inc. cosmology marg. & 10\% & 4.1\% & 8.4\% \\
        \hline
        SKA-2 (SKADS) & 3.9\% & 1.2\% & 0.35\% \\
        inc. cosmology marg. & 9.0\% & 2.2\% & 11.5\% \\
        \hline
    \end{tabular}
    \label{tab:zparm_fisher}
\end{table}

\subsection{Inference on Cosmological Parameters with SKA-2}
\begin{figure}[h!]
    \centering
\includegraphics[width=0.5\textwidth]{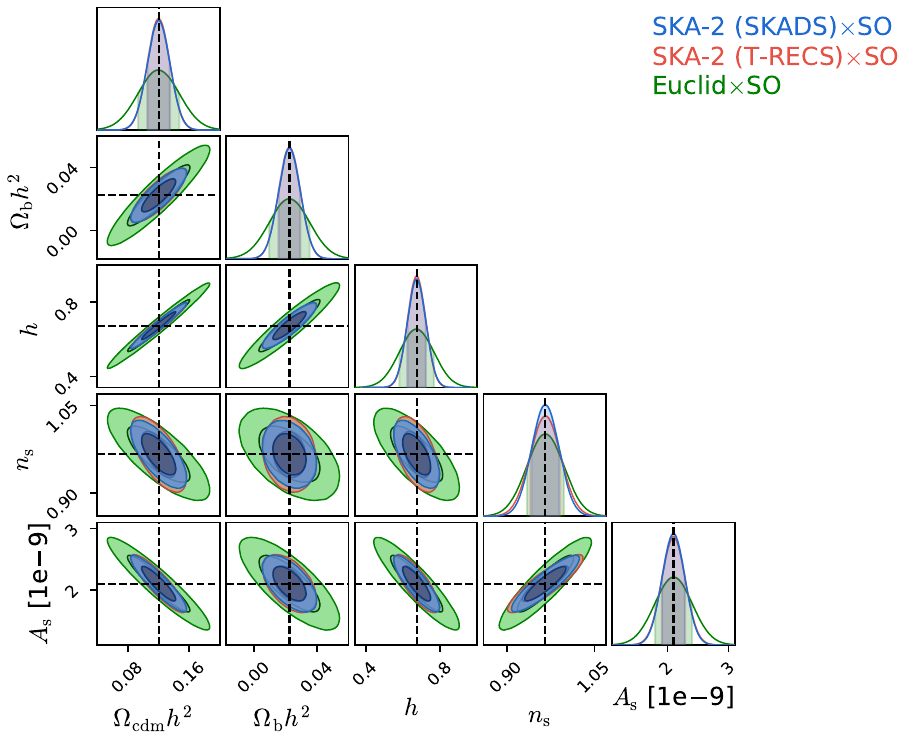}\includegraphics[width=0.5\textwidth]{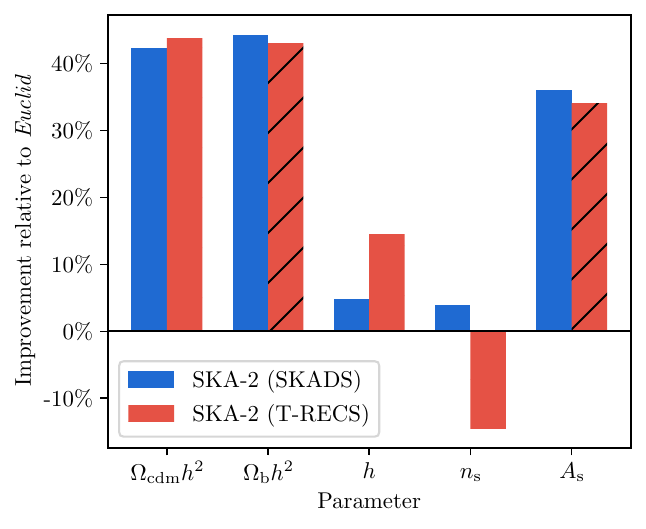}
    \caption{\emph{Left:} Fisher matrix forecasts for cosmological constraining power from the experiments considered. \emph{Right:} Relative improvement on cosmological parameters from the SKA cases including marginalisation over $\beta, \gamma$ redshift nuisance variables, with respect to the \textit{Euclid} case with no redshift nuisance parameters.}
    \label{fig:improvements}
\end{figure}

Next we look at the relative cosmological constraining power of the SKA-2 and \emph{Euclid}-like experiments considered (we do not show cosmological results for SKA-1 as the constraining power is low). The Fisher contours for T-RECS, SKADS and \textit{Euclid} are shown in Fig.~\ref{fig:improvements}. Here, we are specifically interested in the relative cosmological constraining power as we are not considering nuisance parameters (such as shear calibration, non-linear power spectra or Intrinsic Alignments) and are not considering the tomographic analyses that are normally used in cosmic shear surveys to maximise their absolute constraining power. Here we assume that these may be considered to be similar across the radio and optical experiments (for further discussions of how much this may be the case see \cite{brown:ska_weak_lensing, harrison:skaforecasts}). We find that the $\gamma_\mathrm{R}\times \kappa_\mathrm{CMB}$ cross-correlation tightens the constraints on all parameters with respect to $\gamma_\mathrm{O}\times \kappa_\mathrm{CMB}$. The specific $1\sigma$ errors and the relative improvement with respect to \textit{Euclid} are reported in Tab.~\ref{tab:cosmology_fisher} and \cref{fig:improvements}. The overall improvement can be attributed to the long-tail galaxy distribution, which allows to explore higher redshifts compared to optical surveys, and the relatively large sky overlap between SKA-2 and SO. The combination of these two factors reduces the $1\sigma$ errors up to $\mathcal{O}(30\%)$ with respect to \textit{Euclid}. In the constraints shown in \cref{tab:cosmology_fisher} we consider first fixing the redshift parameters as in the case where they are strongly calibrated externally such as through clustering redshifts \cite{morrison:wizz, alonso:clusteringz, cunnington:clusteringz} or another method (rows one and four), and second marginalising over the redshift distribution model parameters where indicated.

\begin{table}
    \renewcommand{\arraystretch}{1.2} 
    \centering
    \caption{The table reports the predicted $1\sigma$ errors on the fiducial $\Lambda$CDM parameters. The first and fourth rows are for fixed redshift parameters and the other rows for the indicated marginalisations. We also show the relative improvement (r.i.) with respect to the case of the \emph{Euclid}-like experiment with fixed redshift parameters. Rows two and five are further represented in \cref{fig:improvements}.}
    \label{tab:model_params}
    \begin{tabular}{lccccc}
        \hline
        Experiment & $\sigma_{\omega_\mathrm{cdm}h^2}$ (r.i.) & $\sigma_{\omega_\mathrm{b}h^2}$ (r.i.) & $\sigma_{h}$ (r.i.) & $\sigma_{n_\mathrm{s}}$ (r.i.) & $\sigma_{10^{9}A_\mathrm{s}}$ (r.i.)\\
        \hline
SKA-2 (SKADS) & 0.014 (48.4\%)	&	0.0066 (48.3\%)	&	0.047 (49.4\%)	&	0.022 (27.7\%)	&	1.8 (41.4\%)	\\
inc. $\beta, \gamma$ marg. & 0.015 (42.3\%)	&	0.0071 (44.3\%)	&	0.088 (4.9\%)	&	0.030 (3.9\%)	&	1.9 (36.0\%)	\\
inc. $z_{\rm tail}$ marg. & 0.014 (46.0\%)	&	0.0068 (46.5\%)	&	0.075 (18.4\%)	&	0.030 (4.4\%)	&	1.8 (40.1\%)	\\
\hline
SKA-2 (T-RECS) & 0.013 (49.8\%)	&	0.0064 (49.4\%)	&	0.045 (51.3\%)	&	0.025 (19.5\%)	&	1.8 (39.7\%)	\\
inc. $\beta, \gamma$ marg. & 0.015 (43.7\%)	&	0.0072 (43.0\%)	&	0.079 (14.4\%)	&	0.036 ($-14.5\%$)	&	2.0 (34.0\%)	\\
inc. $z_{\rm tail}$ marg. & 0.014 (47.6\%)	&	0.0069 (45.7\%)	&	0.070 (23.4\%)	&	0.035 ($-12.3\%$)	&	1.9 (37.2\%)	\\
        \hline
    \end{tabular}
    \label{tab:cosmology_fisher}
\end{table}

\begin{figure}
    \centering
    \includegraphics[width=0.475\linewidth]{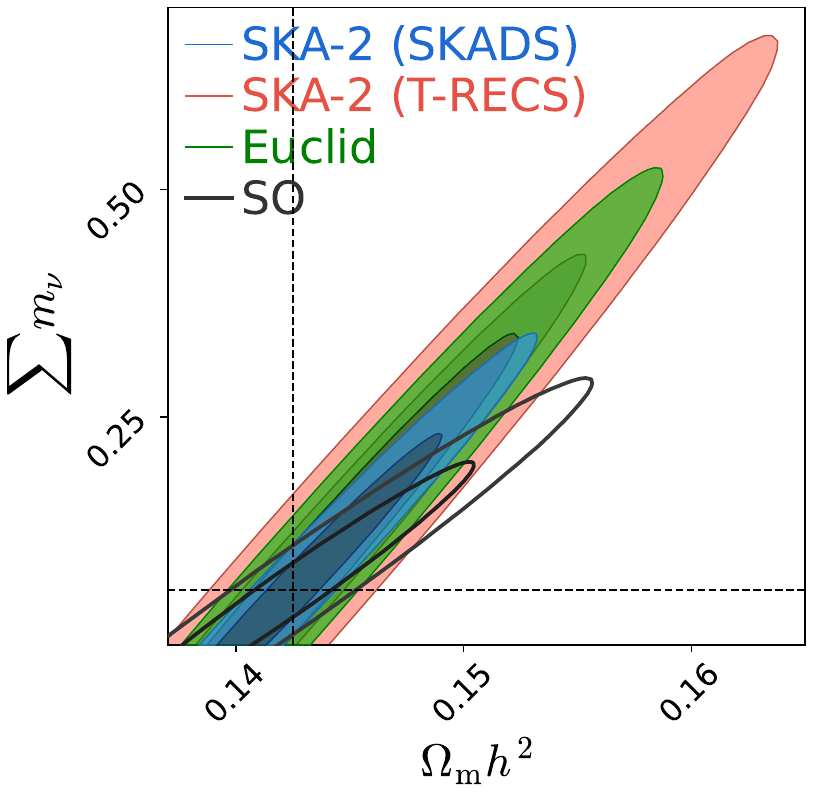}\includegraphics[width=0.475\linewidth]{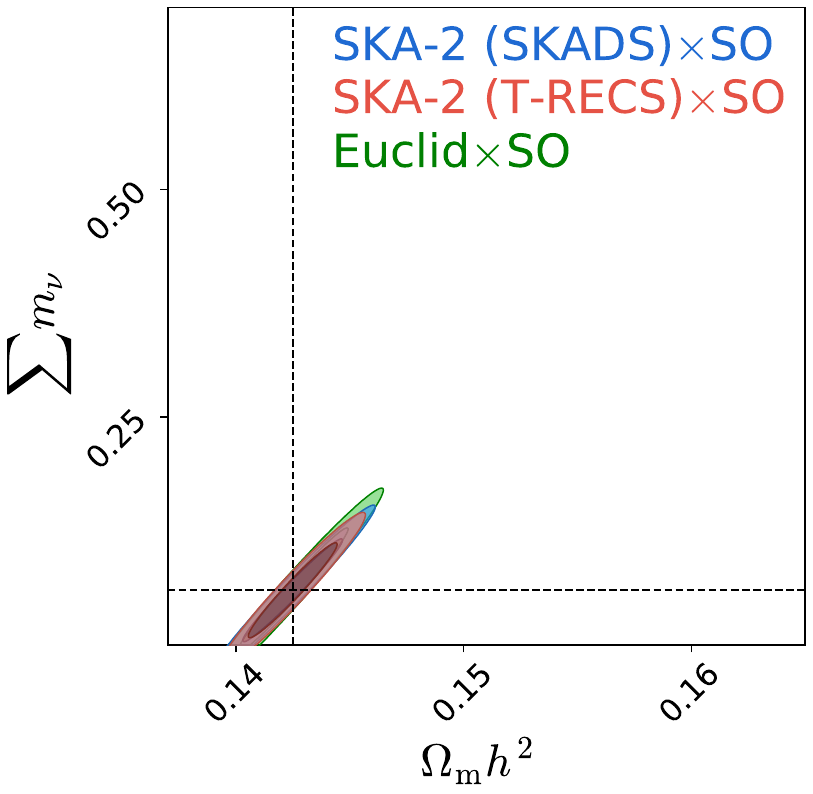} 
    \caption[Fisher prediction of 68\% and 95\% contours on $\{\omega_\mathrm{m},\,\sum m_\nu\}$.]{Fisher prediction of 68\% and 95\% contours on $\omega_\mathrm{m}$ and $\sum m_\nu$ from the full $\gamma\times\kappa_\mathrm{CMB}$ cross-correlations with radio and optical surveys (\textit{right panel}). In the \textit{left panel} we show the constraints from the auto-correlation, including CMB $\kappa$ only.}
        \label{fig:fisher_nu}
\end{figure}

Fixing all other cosmological parameters, we now focus on constraining the parameter space $\{\omega_\mathrm{m},\,\sum m_\nu\}$. Neutrinos contribute to the total matter energy density through 
\begin{equation}
    \Omega_\nu \simeq \frac{\sum m_\nu}{93.14\,\mathrm{eV}\,h^2}.
\end{equation}
In Fig.~\ref{fig:fisher_nu}, we show the Fisher contours and report the $1\sigma$ error values in Tab.~\ref{tab:fisher_nu}. We also show the constraint resulting from $C_\ell^{\gamma\gamma}$, the auto-correlation, to show that the effect of combining auto- and cross-correlated spectra is quite significant for this parameter space. In addition, the different redshift distributions of the SKA-2 experiment both result in improvements compared to the \emph{Euclid}-like experiment, at the level of $23\%$ and $16\%$ for T-RECS and SKADS models respectively.


 \begin{table}[h!]
        \renewcommand{\arraystretch}{1.2} 
        \centering
        \begin{tabular}{l c c}
        \toprule 
        Experiment & $\sigma(\omega_\mathrm{m})$ (r.i.) & $\sigma(\sum m_\nu)$ (r.i.)\\ \midrule 
        SKA-2$\times$SO (T-RECS) & $0.0013$ (18.50\%) &  $0.034$ (22.56\%)\\ \midrule
        SKA-2$\times$SO (SKADS) & $0.0015$ (6.67\%) & $0.038$ (15.58\%) \\  
        \bottomrule
        \end{tabular}
        \caption [Predicted $1\sigma$ errors \{$\omega_\mathrm{m}$, $\sum m_\nu$\}.]{The table reports $1\sigma$ errors on the subset of neutrino parameters $\omega_\mathrm{m}$ and $\sum m_\nu$ for SKA-2 cross-correlations, assuming the different models for galaxy number density. Between brackets we show the relative improvement (\%) with respect to a \textit{Euclid}-like experiment.}
        \label{tab:fisher_nu}
\end{table}

\section{Conclusion}
\label{sec:conc}
In this study, we have explored the cross-correlation between radio cosmic shear and CMB lensing convergence ($\gamma_\mathrm{R}\times \kappa_\mathrm{CMB}$) as a tool to refine constraints on radio source redshift distribution parameters, cosmological parameters of the $\Lambda$CDM model and one of its extensions, namely the sum of neutrino masses. We focused on the SKA's mid-frequency telescope as an example of a future radio continuum telescope and investigated its relative power against a \textit{Euclid}-like optical survey. 

Using two detailed simulations of the continuum radio sky and the Simons Observatory (SO) as a CMB experiment, we found that the full $\gamma_\mathrm{R}\times \kappa_\mathrm{CMB}$ cross-correlation, which include auto- and cross-power spectra, can provide valuable information about the redshift distribution of galaxies, thereby addressing a key challenge in radio weak lensing. By combining these cosmic shear and CMB lensing observables with galaxy clustering of the radio sources we expect to also be able to measure their linear galaxy bias, another highly valuable piece of information which otherwise dilutes cosmological constraining power.

We further show that constraints on cosmological parameter improve with respect to the combination of auto- and cross-correlation between SO and \textit{Euclid}. More specifically, we observed that the cross-correlation can tighten constraints by up to $30\%$ and narrow down the sum of neutrino masses by approximately $24\%$. These results highlight the potential of radio weak lensing in complementing optical surveys and refining our understanding of cosmological parameters, particularly in the neutrino sector.
These results can be attributed mainly to the distribution of sources characterized which extend to higher redshift than the optical counterpart. The large field of view of radio surveys, which allows a large overlap with CMB observations, also affects the results.

This work can be extended in many directions. Firstly, it is important to notice that the constraints obtained in our study are optimistic, as we did not include uncertainties related to various factors such as galaxy intrinsic alignments and non-linear matter clustering modeling. 
Secondly, we did not perform a tomographic analysis, which could, in turn, improve the constraints found here.
Furthermore, while data from SO and CMB Stage-4 (CMB-S4)~\cite{cmbs4:science_book} will provide a precise 2D map of the total matter distribution, precise measurements require an excellent control of observational systematics that hinder the signal. In analogy with optical surveys, the combination of radio and CMB data could eliminate some systematic effects and suppress biases. Finally, the cross-correlation could be extended to higher-order correlation functions, exploring the bispectrum of combinations of $\gamma_\mathrm{R}$ and $\kappa_\mathrm{CMB}$. Apart from adding more information, the bispectrum could break parameter degeneracies.

In conclusion, the cross-correlation between radio cosmic shear and CMB lensing convergence opens up new avenues for exploring cosmological parameters and offers exciting prospects for future cosmological studies.

\section*{Acknowledgements}
This is not an official Simons Observatory Collaboration product. WRC would like to thank the Simons Foundation for support during the initial stages of this project. IH acknowledges support from the European Research Council (ERC) under the European Union's Horizon 2020 research and innovation programme (Grant agreement No. 849169). We thank Giulia Piccirilli for helpful comments on the draft. In addition to the references in the main text we thank the authors and maintainers of public software codes including \texttt{NumPy}~\citep{harris_array_2020}, \texttt{SciPy}~\citep{virtanen_scipy_2020}, \texttt{matplotlib}~\citep{hunter_matplotlib_2007}.

\subsubsection*{Author contributions}
\par We list here the roles and contributions of the authors according to the Contributor Roles Taxonomy (CRediT)\footnote{\url{https://credit.niso.org/}}. \newline
\textbf{Alba Kalaja}: Conceptualization (supporting), Investigation (lead), Methodology (lead), Visualization (equal), Writing - original draft (equal). \textbf{Ian Harrison}: Investigation (supporting), Methodology (supporting), Visualization (equal), Writing - original draft (equal). \textbf{Willian R. Coulton}: Conceptualization (lead), Investigation (supporting), Methodology (supporting), Visualization (supporting), Writing - original draft (supporting).

\bibliographystyle{utcaps}
\bibliography{references}

\providecommand{\href}[2]{#2}\begingroup\raggedright\begin{thebibliography}{10}

\bibitem{lewis:weak_lensing_review}
A.~Lewis and A.~Challinor, ``Weak gravitational lensing of the CMB'', \href{http://dx.doi.org/10.1016/j.physrep.2006.03.002}{{\em Physics Reports} {\bfseries 429} no.~1, (Jun, 2006) 1–65}.

\bibitem{kilbinger:cosmic_shear}
M.~Kilbinger, ``Cosmology with cosmic shear observations: a review'', \href{http://dx.doi.org/10.1088/0034-4885/78/8/086901}{{\em Reports on Progress in Physics} {\bfseries 78} no.~8, (Jul, 2015) 086901}, \href{http://arxiv.org/abs/1411.0115}{{\ttfamily arXiv:1411.0115}}.

\bibitem{Kilbinger_2015}
M.~{Kilbinger}, ``{Cosmology with cosmic shear observations: a review}'', \href{http://dx.doi.org/10.1088/0034-4885/78/8/086901}{{\em Reports on Progress in Physics} {\bfseries 78} no.~8, (July, 2015) 086901}, \href{http://arxiv.org/abs/1411.0115}{{\ttfamily arXiv:1411.0115 [astro-ph.CO]}}.

\bibitem{Bartelmann_2017}
M.~{Bartelmann} and M.~{Maturi}, ``{Weak gravitational lensing}'', \href{http://dx.doi.org/10.4249/scholarpedia.32440}{{\em Scholarpedia} {\bfseries 12} no.~1, (Jan., 2017) 32440}, \href{http://arxiv.org/abs/1612.06535}{{\ttfamily arXiv:1612.06535 [astro-ph.CO]}}.

\bibitem{das:detection_lens_act}
S.~Das {\em et~al.}, ``Detection of the Power Spectrum of Cosmic Microwave Background Lensing by the Atacama Cosmology Telescope'', \href{http://dx.doi.org/10.1103/physrevlett.107.021301}{{\em Phys. Rev. L} {\bfseries 107} no.~2, (Jul, 2011) }.

\bibitem{das:detection_lens_act2}
S.~Das {\em et~al.}, ``The Atacama Cosmology Telescope: temperature and gravitational lensing power spectrum measurements from three seasons of data'', \href{http://dx.doi.org/10.1088/1475-7516/2014/04/014}{{\em JCAP} {\bfseries 2014} no.~04, (Apr, 2014) 014–014}.

\bibitem{planck13:gravitational_lensing}
P.~A.~R. Ade {\em et~al.} \href{http://dx.doi.org/10.1051/0004-6361/201321543}{{\em Astronomy \& Astrophysics} {\bfseries 571} (Oct, 2014) A17}.

\bibitem{ade:polarbear_lens_detection}
P.~A.~R. Ade {\em et~al.}, ``Measurement of the Cosmic Microwave Background Polarization Lensing Power Spectrum with the POLARBEAR Experiment'', \href{http://dx.doi.org/10.1103/physrevlett.113.021301}{{\em Phys. Rev. L} {\bfseries 113} no.~2, (Jul, 2014) }.

\bibitem{spt:lensing_measurement}
A.~van Engelen {\em et~al.} \href{http://dx.doi.org/10.1088/0004-637x/756/2/142}{{\em The Astrophysical Journal} {\bfseries 756} no.~2, (Aug, 2012) 142}.

\bibitem{story:spt_lens_detection2}
K.~T. Story {\em et~al.}, ``A Measurement of the Cosmic Microwave Background Gravitational Lensing Potential from 100 Square Degrees of SPTpol Data'', \href{http://dx.doi.org/10.1088/0004-637x/810/1/50}{{\em The Astrophysical Journal} {\bfseries 810} no.~1, (Aug, 2015) 50}.

\bibitem{bk_lens_detection}
P.~A.~R. Ade {\em et~al.}, ``{BICEP2 / Keck Array VIII: Measurement of gravitational lensing from large-scale B-mode polarization}'', \href{http://dx.doi.org/10.3847/1538-4357/833/2/228}{{\em Astrophys. J.} {\bfseries 833} no.~2, (2016) 228}, \href{http://arxiv.org/abs/1606.01968}{{\ttfamily arXiv:1606.01968 [astro-ph.CO]}}.

\bibitem{planck18:gravitational_lensing}
N.~Aghanim {\em et~al.} \href{http://dx.doi.org/10.1051/0004-6361/201833886}{{\em Astronomy \& Astrophysics} {\bfseries 641} (Sep, 2020) A8}.

\bibitem{qu:act_lensing23}
F.~J. Qu {\em et~al.}, ``The Atacama Cosmology Telescope: A Measurement of the DR6 CMB Lensing Power Spectrum and its Implications for Structure Growth'', 2023.

\bibitem{battye:neutrino_mass_cmb_lensing}
R.~A. Battye and A.~Moss \href{http://dx.doi.org/10.1103/physrevlett.112.051303}{{\em Phys. Rev. L} {\bfseries 112} no.~5, (Feb, 2014) }.

\bibitem{hikage:subaru}
C.~Hikage {\em et~al.} \href{http://dx.doi.org/10.1093/pasj/psz010}{{\em Publications of the Astronomical Society of Japan} {\bfseries 71} no.~2, (Mar, 2019) }.

\bibitem{2022PhRvD.105b3515S}
L.~F. {Secco}, S.~{Samuroff}, {\em et~al.}, ``{Dark Energy Survey Year 3 results: Cosmology from cosmic shear and robustness to modeling uncertainty}'', \href{http://dx.doi.org/10.1103/PhysRevD.105.023515}{{\em \prd} {\bfseries 105} no.~2, (Jan., 2022) 023515}, \href{http://arxiv.org/abs/2105.13544}{{\ttfamily arXiv:2105.13544 [astro-ph.CO]}}.

\bibitem{2021A&A...645A.104A}
M.~{Asgari}, C.-A. {Lin}, B.~{Joachimi}, B.~{Giblin}, C.~{Heymans}, H.~{Hildebrandt}, A.~{Kannawadi}, B.~{St{\"o}lzner}, T.~{Tr{\"o}ster}, J.~L. {van den Busch}, A.~H. {Wright}, M.~{Bilicki}, C.~{Blake}, J.~{de Jong}, A.~{Dvornik}, T.~{Erben}, F.~{Getman}, H.~{Hoekstra}, F.~{K{\"o}hlinger}, K.~{Kuijken}, L.~{Miller}, M.~{Radovich}, P.~{Schneider}, H.~{Shan}, and E.~{Valentijn}, ``{KiDS-1000 cosmology: Cosmic shear constraints and comparison between two point statistics}'', \href{http://dx.doi.org/10.1051/0004-6361/202039070}{{\em \aap} {\bfseries 645} (Jan., 2021) A104}, \href{http://arxiv.org/abs/2007.15633}{{\ttfamily arXiv:2007.15633 [astro-ph.CO]}}.

\bibitem{brown:ska_weak_lensing}
M.~L. Brown {\em et~al.} \href{http://arxiv.org/abs/1501.03828}{{\ttfamily arXiv:1501.03828}}.

\bibitem{2020PASA...37....7S}
{Square Kilometre Array Cosmology Science Working Group}, D.~J. {Bacon}, R.~A. {Battye}, P.~{Bull}, S.~{Camera}, P.~G. {Ferreira}, I.~{Harrison}, D.~{Parkinson}, A.~{Pourtsidou}, M.~G. {Santos}, L.~{Wolz}, {\em et~al.}, ``{Cosmology with Phase 1 of the Square Kilometre Array Red Book 2018: Technical specifications and performance forecasts}'', \href{http://dx.doi.org/10.1017/pasa.2019.51}{{\em \pasa} {\bfseries 37} (Mar., 2020) e007}, \href{http://arxiv.org/abs/1811.02743}{{\ttfamily arXiv:1811.02743 [astro-ph.CO]}}.

\bibitem{condon:radio_gal}
J.~J. {Condon}, ``Radio emission from normal galaxies.'', \href{http://dx.doi.org/10.1146/annurev.aa.30.090192.003043}{{\em Annual Rev. Astron. Astrophys.} {\bfseries 30} (Jan., 1992) 575--611}.

\bibitem{harrison:skaforecasts}
I.~{Harrison}, S.~{Camera}, J.~{Zuntz}, and M.~L. {Brown}, ``{SKA weak lensing - I. Cosmological forecasts and the power of radio-optical cross-correlations}'', \href{http://dx.doi.org/10.1093/mnras/stw2082}{{\em \mnras} {\bfseries 463} no.~4, (Dec., 2016) 3674--3685}, \href{http://arxiv.org/abs/1601.03947}{{\ttfamily arXiv:1601.03947 [astro-ph.CO]}}.

\bibitem{camera:skaforecasts}
S.~{Camera}, I.~{Harrison}, A.~{Bonaldi}, and M.~L. {Brown}, ``{SKA weak lensing - III. Added value of multiwavelength synergies for the mitigation of systematics}'', \href{http://dx.doi.org/10.1093/mnras/stw2688}{{\em \mnras} {\bfseries 464} no.~4, (Feb., 2017) 4747--4760}, \href{http://arxiv.org/abs/1606.03451}{{\ttfamily arXiv:1606.03451 [astro-ph.CO]}}.

\bibitem{2011MNRAS.410.2057B}
M.~L. {Brown} and R.~A. {Battye}, ``{Polarization as an indicator of intrinsic alignment in radio weak lensing}'', \href{http://dx.doi.org/10.1111/j.1365-2966.2010.17583.x}{{\em \mnras} {\bfseries 410} no.~3, (Jan., 2011) 2057--2074}, \href{http://arxiv.org/abs/1005.1926}{{\ttfamily arXiv:1005.1926 [astro-ph.CO]}}.

\bibitem{Whittaker:2015fma}
L.~Whittaker, M.~L. Brown, and R.~A. Battye, ``{Separating weak lensing and intrinsic alignments using radio observations}'', \href{http://dx.doi.org/10.1093/mnras/stv993}{{\em Mon. Not. Roy. Astron. Soc.} {\bfseries 451} no.~1, (2015) 383--399}, \href{http://arxiv.org/abs/1503.00061}{{\ttfamily arXiv:1503.00061 [astro-ph.CO]}}.

\bibitem{vanhaarlem:lofar}
M.~P. van Haarlem {\em et~al.}, ``{LOFAR}: The {LOw}-Frequency {ARray}'', \href{http://dx.doi.org/10.1051/0004-6361/201220873}{{\em Astronomy {\&}amp Astrophysics} {\bfseries 556} (Jul, 2013) A2}.

\bibitem{jonas:meerkat}
J.~L. {Jonas}, ``MeerKAT - The South African Array With Composite Dishes and Wide-Band Single Pixel Feeds'', \href{http://dx.doi.org/10.1109/JPROC.2009.2020713}{{\em IEEE Proceedings} {\bfseries 97} no.~8, (Aug., 2009) 1522--1530}.

\bibitem{johnston:askap}
S.~Johnston {\em et~al.}, ``Science with the Australian Square Kilometre Array Pathfinder'', \href{http://dx.doi.org/10.1071/as07033}{{\em Publications of the Astronomical Society of Australia} {\bfseries 24} no.~4, (2007) 174--188}.

\bibitem{swarup:gmrt}
G.~Swarup, S.~Ananthakrishnan, V.~K. Kapahi, A.~P. Rao, C.~R. Subrahmanya, and V.~K. Kulkarni, ``The Giant Metre-wave Radio Telescope'', {\em Current Science} {\bfseries 60} no.~2, (1991) 95--105.

\bibitem{condon:nvss}
J.~J. {Condon}, W.~D. {Cotton}, E.~W. {Greisen}, Q.~F. {Yin}, R.~A. {Perley}, G.~B. {Taylor}, and J.~J. {Broderick}, ``The NRAO VLA Sky Survey'', \href{http://dx.doi.org/10.1086/300337}{{\em The Astronomical Journal} {\bfseries 115} no.~5, (May, 1998) 1693--1716}.

\bibitem{becker:first}
R.~H. {Becker}, R.~L. {White}, and D.~J. {Helfand}, ``{The FIRST Survey: Faint Images of the Radio Sky at Twenty Centimeters}'', \href{http://dx.doi.org/10.1086/176166}{{\em The Astrophysical Journal} {\bfseries 450} (Sep, 1995) 559}.

\bibitem{intema:gmrt}
H.~T. Intema, P.~Jagannathan, K.~P. Mooley, and D.~A. Frail, ``The {GMRT} 150 {MHz} all-sky radio survey'', \href{http://dx.doi.org/10.1051/0004-6361/201628536}{{\em Astronomy {\&}amp Astrophysics} {\bfseries 598} (Feb, 2017) A78}.

\bibitem{chang_radio_weak_lensing}
T.~C. Chang, A.~Refregier, and D.~J. Helfand \href{http://dx.doi.org/10.1086/425491}{{\em The Astrophysical Journal} {\bfseries 617} no.~2, (Dec, 2004) 794--810}.

\bibitem{demetroullas:ggl}
C.~{Demetroullas} and M.~L. {Brown}, ``{Galaxy-galaxy and galaxy-cluster lensing with the SDSS and FIRST surveys}'', \href{http://dx.doi.org/10.1093/mnras/stx2366}{{\em \mnras} {\bfseries 473} no.~1, (Jan., 2018) 937--952}, \href{http://arxiv.org/abs/1610.03492}{{\ttfamily arXiv:1610.03492 [astro-ph.GA]}}.

\bibitem{demetroullas:sdss-first}
C.~{Demetroullas} and M.~L. {Brown}, ``{Cross-correlation cosmic shear with the SDSS and VLA FIRST surveys}'', \href{http://dx.doi.org/10.1093/mnras/stv2876}{{\em \mnras} {\bfseries 456} no.~3, (Mar., 2016) 3100--3118}, \href{http://arxiv.org/abs/1507.05977}{{\ttfamily arXiv:1507.05977 [astro-ph.CO]}}.

\bibitem{2020MNRAS.495.1706B}
R.~A. {Battye}, M.~L. {Brown}, C.~M. {Casey}, I.~{Harrison}, N.~J. {Jackson}, I.~{Smail}, R.~A. {Watson}, C.~A. {Hales}, S.~M. {Manning}, C.-L. {Hung}, C.~J. {Riseley}, F.~B. {Abdalla}, M.~{Birkinshaw}, C.~{Demetroullas}, S.~{Chapman}, R.~J. {Beswick}, T.~W.~B. {Muxlow}, A.~{Bonaldi}, S.~{Camera}, T.~{Hillier}, S.~T. {Kay}, A.~{Peters}, D.~B. {Sanders}, D.~B. {Thomas}, A.~P. {Thomson}, B.~{Tunbridge}, L.~{Whittaker}, and {SuperCLASS Collaboration}, ``{SuperCLASS - I. The super cluster assisted shear survey: Project overview and data release 1}'', \href{http://dx.doi.org/10.1093/mnras/staa709}{{\em \mnras} {\bfseries 495} no.~2, (June, 2020) 1706--1723}, \href{http://arxiv.org/abs/2003.01734}{{\ttfamily arXiv:2003.01734 [astro-ph.GA]}}.

\bibitem{harrison:super_class}
I.~Harrison {\em et~al.}, ``{SuperCLASS \textendash{} III. Weak lensing from radio and optical observations in Data Release 1}'', \href{http://dx.doi.org/10.1093/mnras/staa696}{{\em Mon. Not. Roy. Astron. Soc.} {\bfseries 495} no.~2, (2020) 1737--1759}, \href{http://arxiv.org/abs/2003.01736}{{\ttfamily arXiv:2003.01736 [astro-ph.CO]}}.

\bibitem{bonaldi:sims}
A.~Bonaldi, I.~Harrison, S.~Camera, and M.~L. Brown, ``{SKA weak lensing\textendash{} II. Simulated performance and survey design considerations}'', \href{http://dx.doi.org/10.1093/mnras/stw2104}{{\em Mon. Not. Roy. Astron. Soc.} {\bfseries 463} no.~4, (2016) 3686--3698}, \href{http://arxiv.org/abs/1601.03948}{{\ttfamily arXiv:1601.03948 [astro-ph.CO]}}.

\bibitem{allison:act_radio_bias}
R.~Allison {\em et~al.}, ``The Atacama Cosmology Telescope: measuring radio galaxy bias through cross-correlation with lensing'', \href{http://dx.doi.org/10.1093/mnras/stv991}{{\em Monthly Notices of the Royal Astronomical Society} {\bfseries 451} no.~1, (Jun, 2015) 849--858}.

\bibitem{namikawa:radio_delensing}
T.~Namikawa, D.~Yamauchi, B.~Sherwin, and R.~Nagata, ``Delensing cosmic microwave background B modes with the Square Kilometre Array Radio Continuum Survey'', \href{http://dx.doi.org/10.1103/physrevd.93.043527}{{\em Physical Review D} {\bfseries 93} no.~4, (Feb, 2016) }.

\bibitem{2015aska.confE..20K}
D.~{Kirk}, F.~B. {Abdalla}, A.~{Benoit-L{\'e}vy}, P.~{Bull}, and B.~{Joachimi}, \href{http://dx.doi.org/10.22323/1.215.0020}{``{Cross correlation surveys with the Square Kilometre Array}'',} in {\em Advancing Astrophysics with the Square Kilometre Array (AASKA14)}, p.~20.
\newblock Apr., 2015.
\newblock \href{http://arxiv.org/abs/1501.03848}{{\ttfamily arXiv:1501.03848 [astro-ph.CO]}}.

\bibitem{lindsay:bias_radio}
S.~N. Lindsay, M.~J. Jarvis, M.~G. Santos, M.~J.~I. Brown, S.~M. Croom, S.~P. Driver, A.~M. Hopkins, J.~Liske, J.~Loveday, P.~Norberg, and A.~S.~G. Robotham, ``Galaxy and Mass Assembly: the evolution of bias in the radio source population to z$\sim$1.5'', \href{http://dx.doi.org/10.1093/mnras/stu354}{{\em Monthly Notices of the Royal Astronomical Society} {\bfseries 440} no.~2, (Mar, 2014) 1527--1541}.

\bibitem{hale:agn_bias_clustering}
C.~L. Hale, M.~J. Jarvis, I.~Delvecchio, P.~W. Hatfield, M.~Novak, V.~Smol{\v{c} }i{\'{c}}, and G.~Zamorani, ``The clustering and bias of radio-selected {AGN} and star-forming galaxies in the {COSMOS} field'', \href{http://dx.doi.org/10.1093/mnras/stx2954}{{\em Monthly Notices of the Royal Astronomical Society} {\bfseries 474} no.~3, (Nov, 2017) 4133--4150}.

\bibitem{siewert:lofar}
T.~M. Siewert {\em et~al.}, ``One- and two-point source statistics from the {LOFAR} Two-metre Sky Survey first data release'', \href{http://dx.doi.org/10.1051/0004-6361/201936592}{{\em Astronomy {\&}amp Astrophysics} {\bfseries 643} (Nov, 2020) A100}.

\bibitem{2020JCAP...12..001S}
E.~{Schaan}, S.~{Ferraro}, and U.~{Seljak}, ``{Photo-z outlier self-calibration in weak lensing surveys}'', \href{http://dx.doi.org/10.1088/1475-7516/2020/12/001}{{\em \jcap} {\bfseries 2020} no.~12, (Dec., 2020) 001}, \href{http://arxiv.org/abs/2007.12795}{{\ttfamily arXiv:2007.12795 [astro-ph.CO]}}.

\bibitem{lesgourgues:neutrinos2}
J.~{Lesgourgues} and S.~{Pastor}, ``Neutrino cosmology and Planck'', \href{http://dx.doi.org/10.1088/1367-2630/16/6/065002}{{\em New Journal of Physics} {\bfseries 16} no.~6, (June, 2014) 065002}.

\bibitem{lesgourgues:neutrinos1}
J.~Lesgourgues, G.~Mangano, G.~Miele, and S.~Pastor, \href{http://dx.doi.org/10.1017/CBO9781139012874}{{\em Neutrino Cosmology}}.
\newblock Cambridge University Press, 2013.

\bibitem{act:neutrino_recent}
M.~S. Madhavacheril {\em et~al.}, ``The Atacama Cosmology Telescope: DR6 Gravitational Lensing Map and Cosmological Parameters'', \href{http://arxiv.org/abs/2304.05203}{{\ttfamily arXiv:2304.05203 [astro-ph.CO]}}.

\bibitem{planck2018:cosmological_parameters}
The {\bfseries Planck Collaboration}, N.~Aghanim {\em et~al.}, ``Planck 2018 results. VI. Cosmological parameters'', \href{http://arxiv.org/abs/1807.06209}{{\ttfamily arXiv:1807.06209}}.

\bibitem{limber:approximation}
D.~N. {Limber}, ``The Analysis of Counts of the Extragalactic Nebulae in Terms of a Fluctuating Density Field.'', \href{http://dx.doi.org/10.1086/145672}{{\em The Astrophysical Journal} {\bfseries 117} (Jan., 1953) 134}.

\bibitem{bonaldi:t_recs}
A.~Bonaldi, M.~Bonato, V.~Galluzzi, I.~Harrison, M.~Massardi, S.~Kay, G.~D. Zotti, and M.~L. Brown, ``The Tiered Radio Extragalactic Continuum Simulation (T-{RECS})'', \href{http://dx.doi.org/10.1093/mnras/sty2603}{{\em Monthly Notices of the Royal Astronomical Society} {\bfseries 482} no.~1, (Sep, 2018) 2--19}.

\bibitem{wilman:skads}
R.~J. Wilman, L.~Miller, M.~J. Jarvis, T.~Mauch, F.~Levrier, F.~B. Abdalla, S.~Rawlings, H.-R. Klckner, D.~Obreschkow, D.~Olteanu, and S.~Young, ``A semi-empirical simulation of the extragalactic radio continuum sky for next generation radio telescopes'', \href{http://dx.doi.org/10.1111/j.1365-2966.2008.13486.x}{{\em Monthly Notices of the Royal Astronomical Society} (Jun, 2008) }.

\bibitem{bird:halofit}
S.~Bird, M.~Viel, and M.~G. Haehnelt, ``Massive neutrinos and the non-linear matter power spectrum'', \href{http://dx.doi.org/10.1111/j.1365-2966.2011.20222.x}{{\em Monthly Notices of the Royal Astronomical Society} {\bfseries 420} no.~3, (Jan, 2012) 2551--2561}.

\bibitem{takahashi:halofit}
R.~Takahashi, M.~Sato, T.~Nishimichi, A.~Taruya, and M.~Oguri, ``{Revising the Halofit Model for the Nonlinear Matter Power Spectrum}'', \href{http://dx.doi.org/10.1088/0004-637X/761/2/152}{{\em Astrophys. J.} {\bfseries 761} (2012) 152}, \href{http://arxiv.org/abs/1208.2701}{{\ttfamily arXiv:1208.2701 [astro-ph.CO]}}.

\bibitem{alonso:crosscorr}
D.~{Alonso}, E.~{Bellini}, C.~{Hale}, M.~J. {Jarvis}, and D.~J. {Schwarz}, ``{Cross-correlating radio continuum surveys and CMB lensing: constraining redshift distributions, galaxy bias, and cosmology}'', \href{http://dx.doi.org/10.1093/mnras/stab046}{{\em \mnras} {\bfseries 502} no.~1, (Mar., 2021) 876--887}, \href{http://arxiv.org/abs/2009.01817}{{\ttfamily arXiv:2009.01817 [astro-ph.CO]}}.

\bibitem{simons:forecasts}
P.~{Ade}, J.~{Aguirre}, Z.~{Ahmed}, S.~{Aiola}, A.~{Ali}, D.~{Alonso}, M.~A. {Alvarez}, K.~{Arnold}, P.~{Ashton}, J.~{Austermann}, and et~al., ``{The Simons Observatory: science goals and forecasts}'', \href{http://dx.doi.org/10.1088/1475-7516/2019/02/056}{{\em \jcap} {\bfseries 2019} no.~2, (Feb., 2019) 056}, \href{http://arxiv.org/abs/1808.07445}{{\ttfamily arXiv:1808.07445 [astro-ph.CO]}}.

\bibitem{hu:joint_gal_lens}
W.~{Hu} and B.~{Jain}, ``{Joint galaxy-lensing observables and the dark energy}'', \href{http://dx.doi.org/10.1103/PhysRevD.70.043009}{{\em \prd} {\bfseries 70} no.~4, (Aug., 2004) 043009}, \href{http://arxiv.org/abs/astro-ph/0312395}{{\ttfamily arXiv:astro-ph/0312395 [astro-ph]}}.

\bibitem{emcee}
D.~{Foreman-Mackey}, D.~W. {Hogg}, D.~{Lang}, and J.~{Goodman}, ``emcee: The MCMC Hammer'', \href{http://dx.doi.org/10.1086/670067}{{\em PASP} {\bfseries 125} (2013) 306--312}, \href{http://arxiv.org/abs/1202.3665}{{\ttfamily 1202.3665}}.

\bibitem{morrison:wizz}
C.~B. {Morrison}, H.~{Hildebrandt}, S.~J. {Schmidt}, I.~K. {Baldry}, M.~{Bilicki}, A.~{Choi}, T.~{Erben}, and P.~{Schneider}, ``{the-wizz: clustering redshift estimation for everyone}'', \href{http://dx.doi.org/10.1093/mnras/stx342}{{\em \mnras} {\bfseries 467} no.~3, (May, 2017) 3576--3589}, \href{http://arxiv.org/abs/1609.09085}{{\ttfamily arXiv:1609.09085 [astro-ph.CO]}}.

\bibitem{alonso:clusteringz}
D.~{Alonso}, P.~G. {Ferreira}, M.~J. {Jarvis}, and K.~{Moodley}, ``{Calibrating photometric redshifts with intensity mapping observations}'', \href{http://dx.doi.org/10.1103/PhysRevD.96.043515}{{\em \prd} {\bfseries 96} no.~4, (Aug., 2017) 043515}, \href{http://arxiv.org/abs/1704.01941}{{\ttfamily arXiv:1704.01941 [astro-ph.CO]}}.

\bibitem{cunnington:clusteringz}
S.~{Cunnington}, I.~{Harrison}, A.~{Pourtsidou}, and D.~{Bacon}, ``{H I intensity mapping for clustering-based redshift estimation}'', \href{http://dx.doi.org/10.1093/mnras/sty2928}{{\em \mnras} {\bfseries 482} no.~3, (Jan., 2019) 3341--3355}, \href{http://arxiv.org/abs/1805.04498}{{\ttfamily arXiv:1805.04498 [astro-ph.CO]}}.

\bibitem{cmbs4:science_book}
The {\bfseries CMB-S4 Collaboration}, K.~N. Abazajian {\em et~al.}, ``{CMB-S4 Science Book, First Edition}'', \href{http://arxiv.org/abs/1610.02743}{{\ttfamily arXiv:1610.02743 [astro-ph.CO]}}.

\bibitem{harris_array_2020}
C.~R. {Harris}, K.~J. {Millman}, S.~J. {van der Walt}, R.~{Gommers}, P.~{Virtanen}, D.~{Cournapeau}, E.~{Wieser}, J.~{Taylor}, S.~{Berg}, N.~J. {Smith}, R.~{Kern}, M.~{Picus}, S.~{Hoyer}, M.~H. {van Kerkwijk}, M.~{Brett}, A.~{Haldane}, J.~F. {del R{\'\i}o}, M.~{Wiebe}, P.~{Peterson}, P.~{G{\'e}rard-Marchant}, K.~{Sheppard}, T.~{Reddy}, W.~{Weckesser}, H.~{Abbasi}, C.~{Gohlke}, and T.~E. {Oliphant}, ``{Array programming with NumPy}'', \href{http://dx.doi.org/10.1038/s41586-020-2649-2}{{\em \nat} {\bfseries 585} no.~7825, (Sept., 2020) 357--362}, \href{http://arxiv.org/abs/2006.10256}{{\ttfamily arXiv:2006.10256 [cs.MS]}}.

\bibitem{virtanen_scipy_2020}
P.~Virtanen {\em et~al.}, ``{{SciPy} 1.0: Fundamental Algorithms for Scientific Computing in Python}'', \href{http://dx.doi.org/10.1038/s41592-019-0686-2}{{\em Nature Methods} {\bfseries 17} (2020) 261--272}.

\bibitem{hunter_matplotlib_2007}
J.~D. {Hunter}, ``{Matplotlib: A 2D Graphics Environment}'', \href{http://dx.doi.org/10.1109/MCSE.2007.55}{{\em Computing in Science and Engineering} {\bfseries 9} no.~3, (May, 2007) 90--95}.

\end{thebibliography}\endgroup
\end{document}